\documentstyle[aps,preprint,epsf]{revtex}
\begin{document}

\title{Particle-Number Reprojection in the Shell Model Monte Carlo Method:
Application  to Nuclear Level Densities}

\author{ Y. Alhassid$^1$, S. Liu$^1$ and H. Nakada$^2$}

\address{$^1$Center for Theoretical Physics,
Yale University, New Haven, Connecticut  06520, U.S.A. \\
$^2$Department of Physics, Chiba University,
Yayoi-cho, Chiba 263-8522, Japan}

\date{August 16, 1999}

\maketitle

\begin{abstract}

  We introduce a particle-number reprojection method in the shell model Monte
Carlo  that enables the calculation of observables for a series of
nuclei  using a Monte Carlo sampling  for a single nucleus. The method is
used to calculate nuclear level densities in the complete $(pf+g_{9/2})$-shell
using  a good-sign Hamiltonian. Level densities of odd-$A$ and odd-odd nuclei
are  reliably extracted despite an additional sign problem.
Both the mass and the $T_z$ dependence of the experimental level
densities are well described without any adjustable parameters.
The single-particle level density parameter is found to vary
smoothly  with mass. The odd-even staggering observed in the calculated
backshift  parameter follows the experimental data more closely than do
empirical  formulae.
\end{abstract}


  The interacting shell model has successfully described a variety of nuclear
properties.  However, the size of the model space increases rapidly with the
number  of valence nucleons and/or orbits, and exact diagonalization of the
nuclear  Hamiltonian in a full major shell is limited to nuclei with $A \alt
50$  \cite{fp9,fp11}. The development of quantum Monte Carlo methods for the
nuclear  shell model allowed realistic  calculations of finite- and
zero-temperature  observables in model spaces that are much larger than those
treated  by conventional diagonalization techniques \cite{SMMC,ADK94}.

The Monte Carlo method was successfully adapted to the microscopic
calculations  of nuclear level densities \cite{NA97}. Accurate level
densities are needed for estimating nuclear reaction rates, e.g., neutron and
proton  capture  rates.  The nucleosynthesis of many of the heavy elements
proceeds  by radiative  capture of neutrons ($s$ and $r$ processes) or
protons
($rp$  process) in competition with beta decay \cite{BBF57,RTK97}.
Theoretical  approaches to level  densities are often based on the Fermi gas
model,  i.e.,  the Bethe formula  \cite{Bethe37}, which describes the
many-particle  level density in terms  of the single-particle level density
parameter  $a$.  Shell corrections and  two-body correlations are taken into
account  empirically by defining a fictitious  ground state energy. In the
backshifted  Bethe formula (BBF) the ground  state energy is shifted by an
amount  $\Delta$. This formula describes well  the experimental level
densities  of many nuclei if both $a$ and  $\Delta$ are fitted for each
individual  nucleus \cite{Dilg73}. While these parameters have been discussed
in  terms of their global systematics,
it is difficult to predict their values for particular nuclei.

The nuclear shell model offers an attractive framework to calculate level
densities,  but the model space required to calculate level densities at
excitation  energies in the neutron resonance regime is usually too large for
conventional  diagonalization methods.   We have recently developed a method
\cite{NA97}  to calculate exact level densities using the shell model Monte
Carlo  (SMMC) approach, and applied it to calculate the level densities of
even-even  nuclei from iron to germanium \cite{NA98}.  Fermionic Monte Carlo
methods  are usually hampered by the so-called sign problem, which causes a
breakdown of the method at low temperatures. A practical solution in the
nuclear  case was developed in Ref. \cite{ADK94} but the associated
extrapolation  errors are too large for reliable calculations of level
densities.  In Ref. \cite{NA97} the sign problem was overcome by constructing
a  good-sign interaction in the $(pf + g_{9/2})$-shell that includes
correctly the  dominating collective components of realistic effective
interactions  \cite{DZ96}.   The SMMC level densities are well-fitted by the
BBF,  and both $a$  and $\Delta$ can be extracted from the microscopic
calculations.

  The SMMC approach is computationally intensive.
In particular, the  SMMC level densities require calculations of  the thermal
energy  at all temperatures. The weight function used in the random walk is
temperature  dependent, and a new Monte Carlo sampling is required at each
temperature.  Since this procedure has to be repeated for each nucleus, the
calculations  are time-consuming. In this paper we describe a
particle-reprojection  method that allows us to calculate observables
for  a series of nuclei using Monte Carlo sampling for one nucleus only. The
random  walk is done with a weight function proportional to the partition
function  of a given even-even nucleus (which is positive definite for a
good-sign  interaction), and the thermal observables are then calculated for
several  nuclei by reprojection on different particle numbers (both even and
odd).  This method allows significantly more economical calculations of level
densities.  We apply the method in the full $(pf + g_{9/2})$-shell to study
the  systematics of $a$ and  $\Delta$ for even-even, odd-$A$ and odd-odd
manganese,  iron and cobalt nuclei. A direct comparison with both
experimental data and empirical formulae is presented. The agreement with the
data  is remarkably good with no adjustable parameters in the microscopic
calculations.  Furthermore, we find that the SMMC values follow
the  data more closely than do the empirical values.

The Monte Carlo method is based on the
Hubbard-Stratonovich  representation of the imaginary-time propagator,
$e^{-\beta  H} = \int D[\sigma] G(\sigma) U_\sigma$, where $G(\sigma)$ is a
Gaussian  weight and $U_\sigma$ is the propagator of non-interacting
nucleons  moving in fluctuating auxiliary fields $\sigma$ which depend
on imaginary time. The canonical
thermal  expectation value of an observable $O$ is given by $\langle O
\rangle_{\cal  A} = \int D[\sigma] G(\sigma){\rm Tr}_{\cal A}(O
U_\sigma)/\int
D[\sigma]  G(\sigma){\rm Tr}_{\cal A} U_\sigma$, where ${\rm Tr}_{\cal A}$
denotes  a trace in the subspace of a fixed number of particles ${\cal A}$.
In actual calculations we project on both  neutron number $N$ and proton
number
$Z$, and in the following ${\cal A}$ will denote  $(N,Z)$. We
rewrite
\begin{equation}\label{observable}
\langle O \rangle_{\cal A} = \left\langle {{\rm Tr}_{\cal A} (O U_\sigma) /
{\rm  Tr}_{\cal A} U_\sigma} \right\rangle_W \;,
\end{equation}
where we have introduced the notation
\begin{equation}\label{weight-W}
\langle X_\sigma \rangle_W \equiv {\int D[\sigma] W(\sigma) X_\sigma \over
\int  D[\sigma] W(\sigma)} \;,
\end{equation}
and $W(\sigma)\equiv G(\sigma) {\rm Tr}_{\cal A} U_\sigma$.
For an even number of particles with a good-sign interaction, $W(\sigma)$ is
positive definite.   In the Monte Carlo method we choose $M$ samples
(each denoted by $\sigma_k$)  according to the weight function $W(\sigma)$,
and estimate $\langle  X_\sigma \rangle_W \approx \sum_k X_{\sigma_k} /M$.

  We assume that the Monte Carlo sampling is done for a nucleus with particle
number  ${\cal A}$,  and consider the ratio $Z_{{\cal A}^\prime}/Z_{\cal A}$
between  the partition function of a nucleus with
${\cal A}'$ particles and the partition function of the original
nucleus. In the notation of Eq. (\ref{weight-W})
\begin{eqnarray}\label{partition-ratio}
{Z_{{\cal A}^\prime}(\beta) \over Z_{\cal A}(\beta)} \equiv
{{\rm Tr}_{{\cal A}^\prime} e^{-\beta H} \over {\rm Tr}_{\cal A} e^{-\beta
H}}  =
\left\langle { {\rm Tr}_{{\cal A}^\prime} U_\sigma \over
{\rm Tr}_{\cal A} U_\sigma} \right\rangle_W \;.
\end{eqnarray}
Similarly, the expectation value of an observable $O$ for the nucleus with
${\cal
A}'$  particles can be calculated from
\begin{equation}\label{observable'}
\langle O \rangle_{{\cal A}'}= {\left\langle \left({{\rm Tr}_{{\cal A}'}
OU_\sigma  \over {\rm Tr}_{{\cal A}'} U_\sigma}\right) \left({ {\rm
Tr}_{{\cal
A}^\prime}  U_\sigma \over {\rm Tr}_{\cal A} U_\sigma}\right) \right\rangle_W
\over
\left\langle { {\rm Tr}_{{\cal A}'} U_\sigma  \over
{\rm Tr}_{\cal A} U_\sigma} \right\rangle_W
} \;.
\end{equation}
The Monte Carlo walk is carried out by projection on a fixed ${\cal A}$, and
Eqs.  (\ref{partition-ratio}) and (\ref{observable'}) are then used to
calculate  the partition functions and observables for a family of nuclei
with
   ${\cal A}' \neq {\cal A}$.

We applied the method to nuclei in the $(pf+ g_{9/2})$-shell, using the
Hamiltonian  of Ref. \cite{NA97}.  The single-particle energies are computed
in  a central Woods-Saxon potential $V(r)$ plus spin-orbit interaction, while
the  two-body interaction includes a monopole isovector pairing of strength
$g_0$   plus a separable surface-peaked interaction \cite{ABDK} $v(\bbox{r},
\bbox{r}^\prime)
= -\chi (dV/dr)(dV/dr^\prime)  \delta(\hat{\bbox{r}} -
\hat{\bbox{r}}^\prime)$.  The surface-peaked interaction is expanded into
multipoles  and only the quadrupole, octupole and hexadecupole terms are
kept.
The strength $\chi$ is determined self-consistently and renormalized. The
strength  of the pairing interaction $g_0 \approx 0.2$ is determined from
experimental   odd-even mass differences.
Both the pairing and the surface-peaked
interactions  are attractive and lead to a good-sign Hamiltonian. A repulsive
isospin-dependent  interaction leads to a sign problem, and was included
perturbatively   in recent level density calculations in  $sd$-shell
\cite{Or97}   and $pf$-shell\cite{La98} nuclei.

In the particle-reprojection method described above we have assumed that the
Hamiltonian   $H$ is independent of ${\cal A}$. Suitable corrections should
be
made   if some of the Hamiltonian parameters vary with ${\cal A}$.  Since
$\chi$   depends only weakly on the mass number $A$ ($\propto A^{-1/3}$), and
the   pairing strength $g_0$ is constant through the shell, the largest
variation  is that of the single-particle energies of the orbit $a$,
$\epsilon_a({\cal A})$. To
correct this variation we approximate the thermal energy of ${\cal
A}'\equiv(N',Z')$  particles  by
\begin{equation}\label{energy}
E_{{\cal A}'} (\beta) \approx \sum_a [\epsilon_a({\cal A}') -\epsilon_a({\cal
A})]\langle  n_a \rangle_{{\cal A}'} + \langle H \rangle_{{\cal A}'}
\end{equation}
where $H$ is the Hamiltonian for a nucleus with ${\cal A}$ particles. In
calculating  the energy for ${\cal A}'$ particles from (\ref{energy}), we
used in  the propagator ($e^{-\beta H}$) the Hamiltonian $H$ for nucleus
${\cal  A}$ rather  than ${\cal A}'$. To minimize the error we reproject on
nuclei  with $N'-Z'$  values close to $N - Z$ (the Woods-Saxon potential
depends  on $N-Z$). In  the applications below we have checked that the
resulting  error in the level density is negligible.

  We used the reprojection method to calculate the thermal energies versus
$\beta$  for $^{50-56}$Mn, $^{52-58}$Fe, and $^{54-60}$Co including odd-$A$
and  odd-odd nuclei.  We sampled according to the even-even nucleus $^{56}$Fe
and  reprojected onto $^{53-56}$Mn, $^{54-58}$Fe, and $^{54-60}$Co, while the
nuclei  $^{50-52}$Mn and $^{52,53}$Fe are reprojected from Monte Carlo
sampling  of the odd-odd $N=Z$ nucleus $^{54}$Co. The calculations are done
for  values of $\beta$ between $\beta=0$ and 1 MeV$^{-1}$ in steps of $\Delta
\beta=1/16$,  and between 1 and 2.5 in steps of $\Delta \beta=1/8$. At each
$\beta$  we used about 4000 independent samples.  Reprojected energy
calculations  typically have larger statistical errors at larger values of
$\beta$.  Therefore we also performed direct Monte Carlo runs (without
reprojection)  for the above nuclei at several values of $\beta$ between
$1.75$  and $3.0$ MeV$^{-1}$. For odd-$A$ and odd-odd nuclei, a typical
statistical  error for the energy at $\beta \sim 2.5$ is $\sim 0.5$
MeV,  while for $\beta \agt 3$ the error is too
large  for the data to be useful. This is just another manifestation of the
sign  problem for nuclei with an odd number of protons and/or neutrons.
Fortunately, because of the high degeneracy in the vicinity of the ground
state  of these nuclei, the  thermal energy is already close to its
asymptotic value.

\begin{figure}

\vspace{1 cm}

\centerline{\epsffile{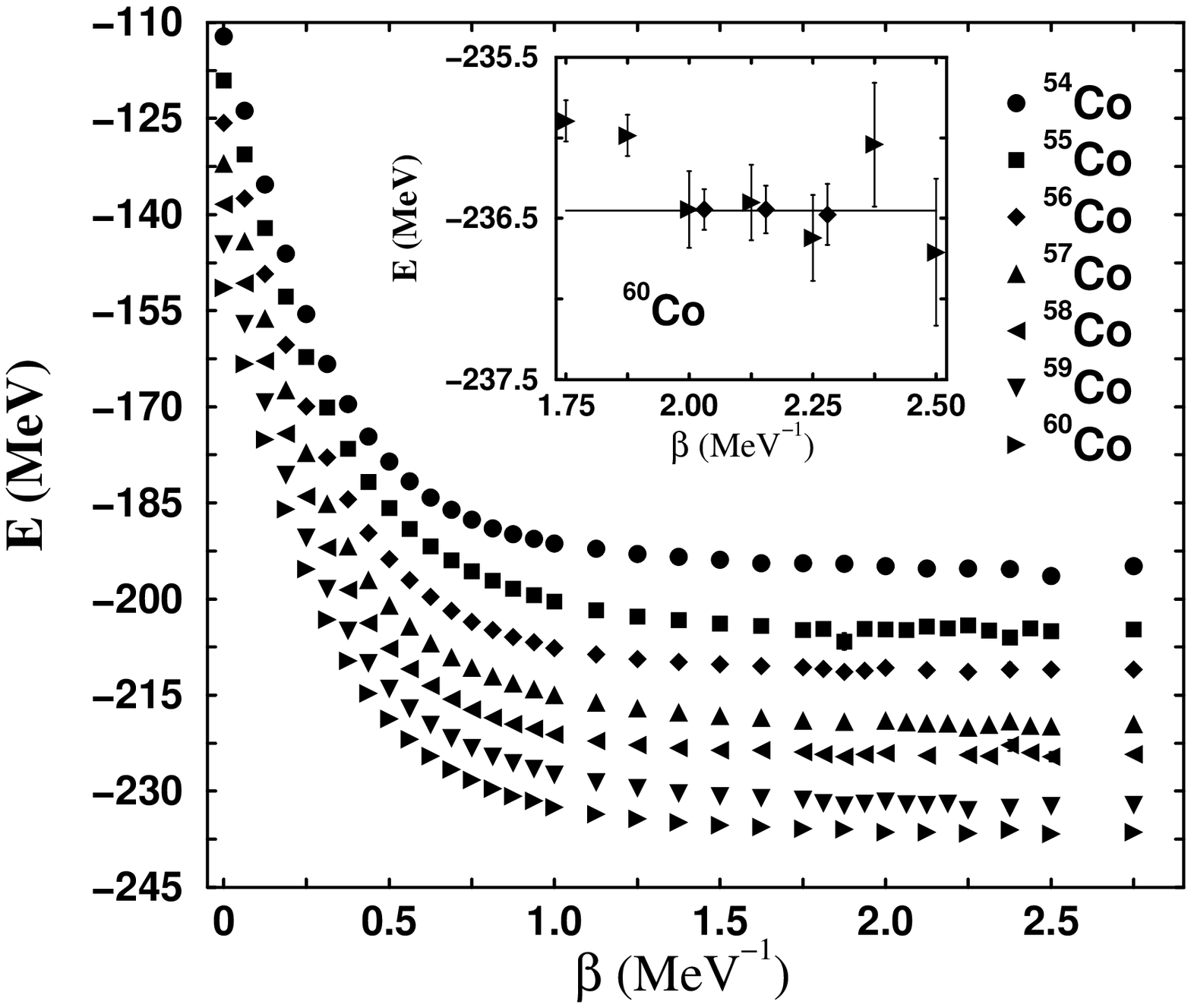}}

\vspace{1 cm}

\caption { The SMMC thermal energies versus $\beta$ for
$^{54-60}$Co (symbols).  Shown on the right are the extrapolated
values of the ground-state energy.  Inset: the SMMC thermal
energies (triangles) at large $\beta$ values for  $^{60}$Co.
The diamonds are the energies obtained by averaging the
large-$\beta$ results above  the corresponding $\beta$. }
\label{fig:Co-e}

\vspace{3 cm}

 \centerline{\epsffile{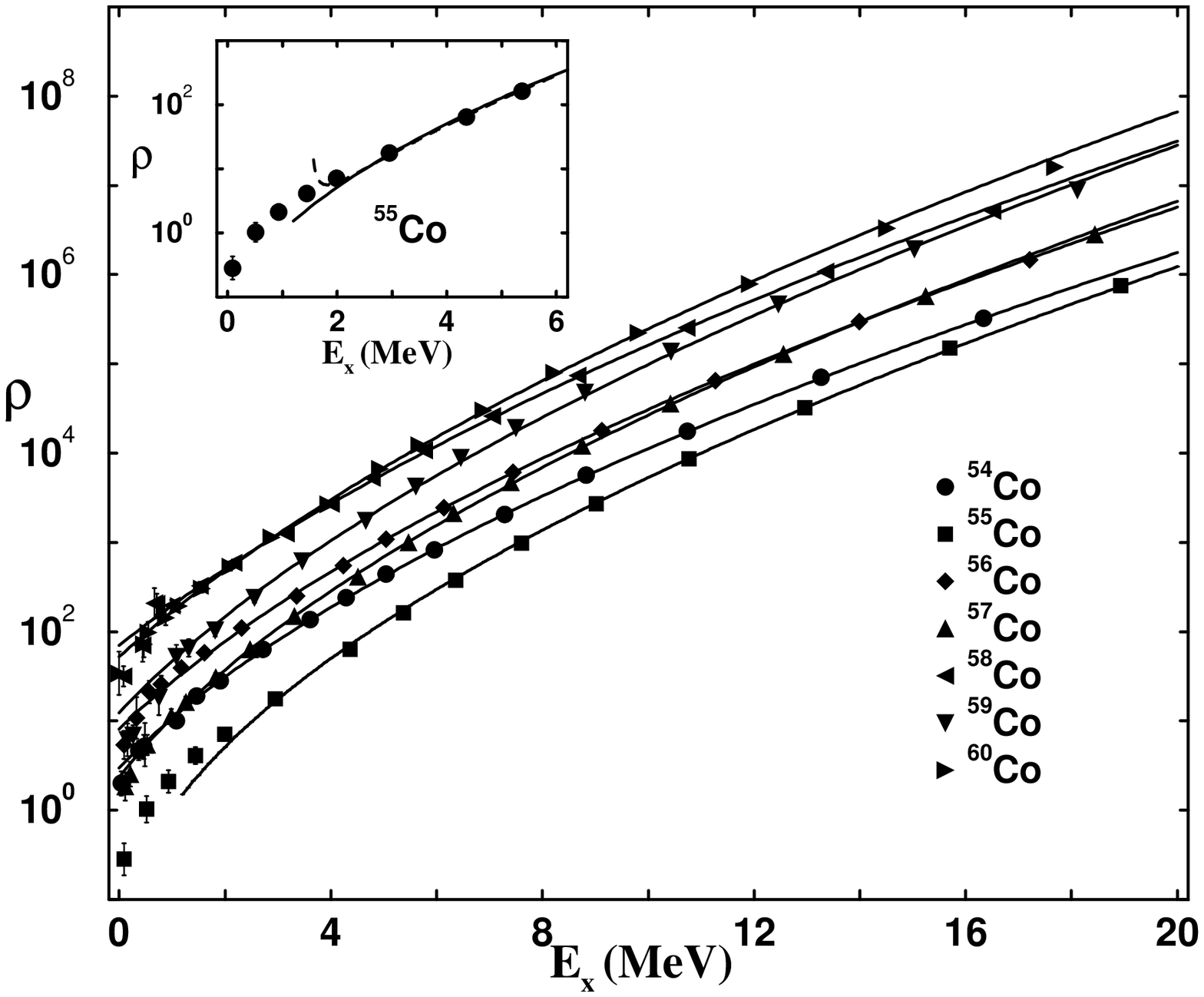}}

\vspace{1 cm}

\caption
{ The SMMC level densities of the $^{54-60}$Co isotopes. The solid lines
describe  a fit to the BBF (\protect\ref{BBF}). The top inset shows the level
density  of $^{55}$Co at low excitation energies. The circles are the SMMC
results,  the solid line is a fit to (\protect\ref{BBF}), and the dashed line
is  a fit to (\protect\ref{BBF}) but without the $t$ term.
}
\label{fig:Co-lvl}

\vspace{1 cm}

 \centerline{\epsffile{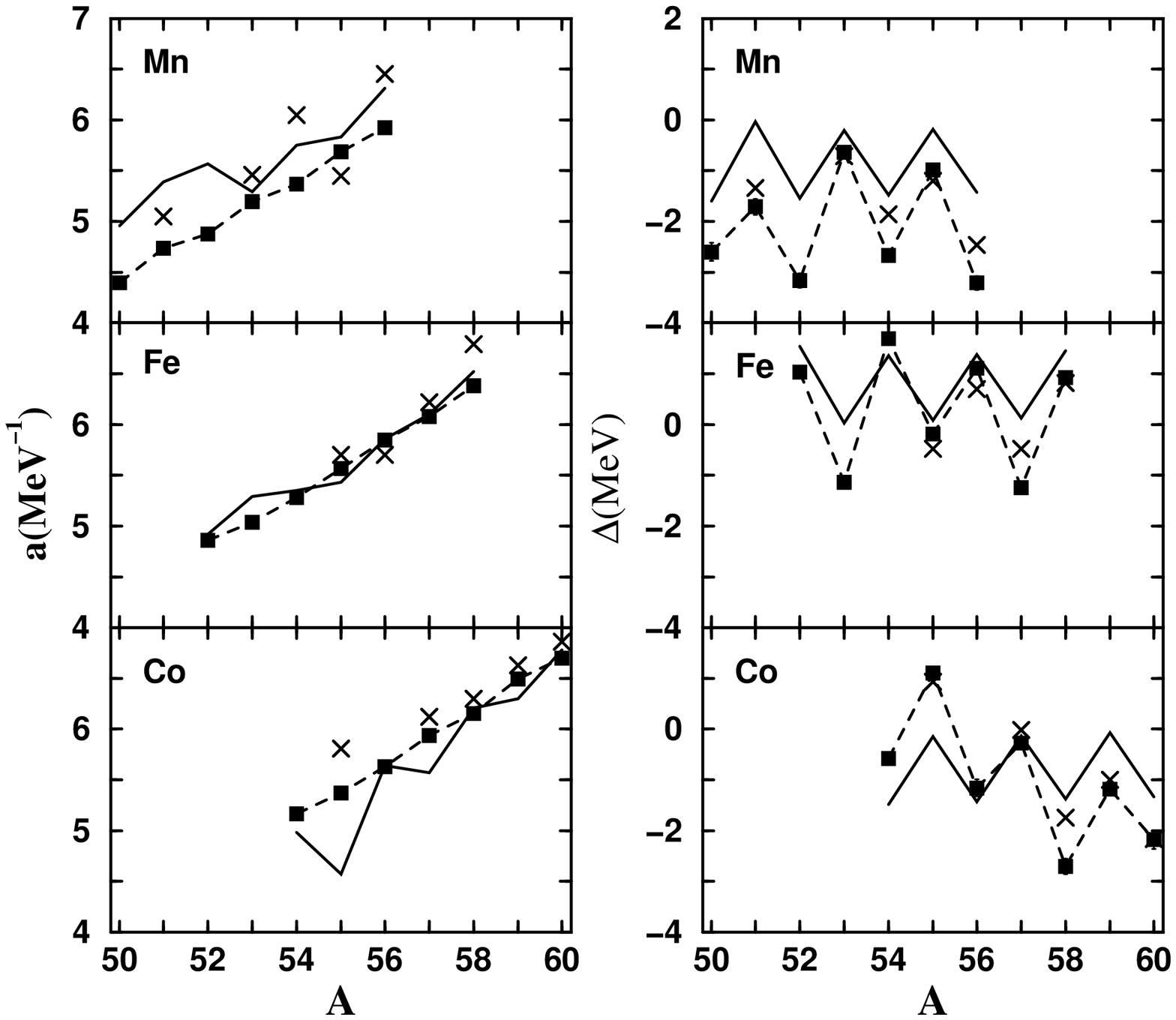}}

\vspace{1 cm}

\caption
{The single-particle level density parameter $a$ (left column) and the
backshift  parameter $\Delta$ (right column) for Mn, Fe and Co isotopes. The
solid  squares (connected by dashed lines) are the results of fitting the
calculated  SMMC level densities to Eq. (\protect\ref{BBF}), while the X's
are
the  experimental results. All the experimental values are from the
compilations  of \protect\cite{Dilg73} (assuming rigid body moment of
inertia),  except for $^{58}$Co and $^{59}$Co where the values quoted in
\protect\cite{Gr74}  are used.  The solid lines are the empirical formulae of
\protect\cite{HWFZ}.
}
\label{fig:a-delta}

\vspace{3 cm}

 \centerline{\epsffile{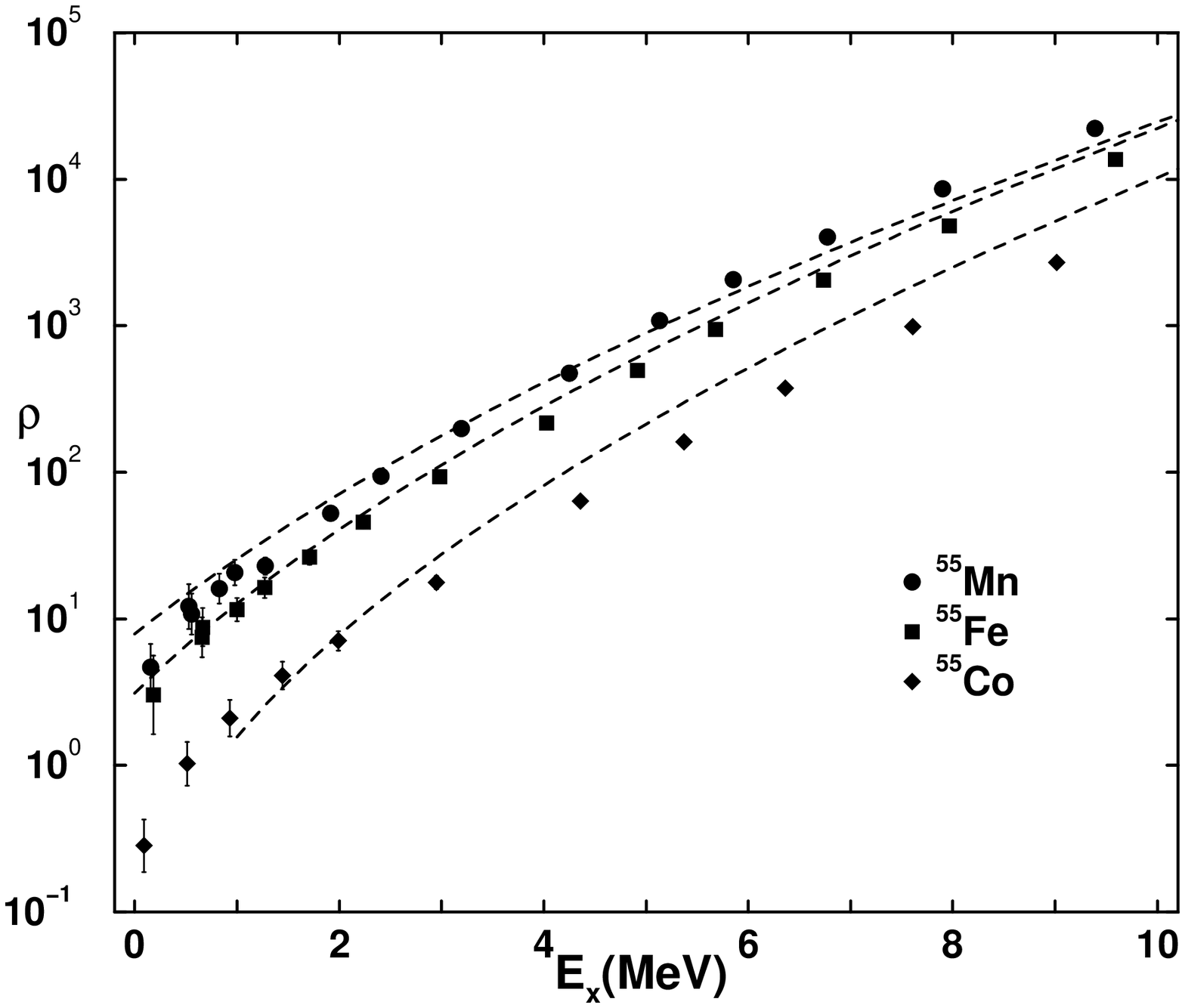}}

\vspace{1 cm}

\caption
{The SMMC level densities of three $A=55$ nuclei: $^{55}$Mn (circles),
$^{55}$Fe  (squares) and $^{55}$Co (diamonds). The dashed lines are the
experimental  level densities.
}
\label{fig:a55-lvl}

\end{figure}

Fig. \ref{fig:Co-e} shows the calculated SMMC thermal energies versus $\beta$
for  a series of cobalt isotopes. The effect of pairing on the thermal
energies  at low temperatures (i.e. large $\beta$) is clearly seen in their
uneven  spacings. The inset of Fig. \ref{fig:Co-e} shows the SMMC thermal
energies (triangles with error bars) for $^{60}$Co for the large values of
$\beta$ only.

In calculating the level density versus excitation energy, it is  important
to  get accurate values of the ground state energy. In Ref. \cite{NA98} we
used,  for even-even nuclei, a two-state model ($0^+$ and $2^+$) to obtain a
two-parameter  fit to the thermal energy and $\langle\bbox J^2 \rangle$. For
odd-$A$  and odd-odd nuclei this method is not useful since in general we do
not  know  the spin of the ground state and first excited state. Moreover,
these  nuclei do not have a gap and often more than two levels contribute to
the  thermal energy at the lowest temperatures for which Monte Carlo
calculations  are still possible.  We estimate the ground state energy of
these  nuclei by taking an average of the large-$\beta$ SMMC values for the
thermal  energy. The diamonds in the inset of Fig. \ref{fig:Co-e} are such
average  values for $^{60}$Co. We estimate the ground state energy of the
odd-$A$  and odd-odd nuclei to be reliable to about $\sim 0.3$ MeV.

To calculate the level density we first find the partition function
$Z_{{\cal  A}'}$  by integrating the relation $-\partial \ln Z_{{\cal
A}'}/\partial  \beta = E_{{\cal A}'}$. The level density is then given by
\begin{eqnarray}\label{level}
  \rho_{{\cal A}'} = (2\pi \beta^{-2} C_{{\cal A}'})^{-1/2}  e^{S_{{\cal
A}'}}  \;,
\end{eqnarray}
  in terms of the canonical entropy
$S_{{\cal A}'} = \beta E_{{\cal A}'} +  \ln Z_{{\cal A}'} $ and the heat
capacity
$C_{{\cal A}'}  = - \beta^2 dE_{{\cal A}'} / d\beta$.

The level densities for the cobalt isotopes of Fig. \ref{fig:Co-e} are shown
in  Fig. \ref{fig:Co-lvl} as a function of excitation energy.
These total level densities are fitted to
\begin{eqnarray}\label{BBF}
\rho (E_x) \approx
g  {{\sqrt\pi}\over{24}} a^{-\frac{1}{4}} (E_x - \Delta +t)^{-\frac{5}{4}}
e^{2\sqrt{a (E_x - \Delta)}} \;,
\end{eqnarray}
where $t$ is a thermodynamic temperature defined by
$E_x - \Delta = a t^2 - t$ and $g=2$.
Eq. (\ref{BBF}) is a modified version of the BBF derived by Lang and Le
Couteur  \cite{LLC54}. It differs from the usual BBF by the additional
``temperature''  term $t$ in the pre-exponential factor, and provides a
better fit  to the calculated level density at lower excitation energies.
 The solid lines  in Fig. \ref{fig:Co-lvl} are the BBF
level densities (\ref{BBF})
fitted to  the SMMC level densities in the range $E_x < 20$ MeV.
 In general we obtain  a good fit down to energies of $\sim 1$ MeV
or smaller. The inset
shows  the low energy fit for $^{55}$Co. The dashed line is a fit to the BBF
without $t$; this approximation starts to diverge around $2$ MeV due to
the singularity  of the pre-exponential factor $(E_x - \Delta)^{-5/4}$. Notice
that the level density for an odd-odd cobalt  (e.g., $^{54}$Co) is higher
than  the level density of the subsequent odd-even cobalt (e.g., $^{55}$Co)
even  though the latter has a larger mass. This is due to reduced pairing
correlations  in the odd-odd nucleus that lead to a smaller backshift
$\Delta$.

  We extracted the level density parameters $a$ and $\Delta$ for the above
nuclei   by fitting Eq. (\ref{BBF}) to the SMMC level densities.  The results
for  $a$ and $\Delta$ versus mass number $A$ are shown in Fig.
\ref{fig:a-delta}.  The Monte Carlo results (solid squares) are compared with
the  experimental data (X's) quoted in Refs. \cite{Dilg73} and \cite{Gr74}.
The  solid lines describe the results of the empirical formulae of Refs.
\cite{HWFZ}.  The calculated values of $a$ depend smoothly on the mass,
unlike some of the empirical results, and in the case of the cobalt
isotopes  follow the data more closely. The staggering seen in the behavior
of $\Delta$  versus $A$ is a result of pairing effects. In the empirical
formulae,  $\Delta \sim 0$ for odd-even nuclei, is positive for even-even
nuclei  and is negative for odd-odd nuclei. We see that the present values of
$\Delta$ follow rather closely the experimental values, and are in general
in better agreement than the empirical values.
The lower values of $a$ (relative to the experimental values) for the odd-odd
manganese  nuclei are compensated by corresponding
lower values of $\Delta$, and thus do not cause significant discrepancies
in the level densities for $E_x \alt 10$~MeV.

  To demonstrate how the Monte Carlo results improve over the empirical
formulae  we show in Fig. \ref{fig:a55-lvl} the calculated level densities of
three  $A=55$ nuclei (Mn, Fe and Co). According to the empirical formula,
$\Delta  \sim 0$ for odd-A nuclei, and the values of $a$ are similar (since
$A$  is the same). The empirical formulae therefore predict similar level
densities  for these nuclei. However the SMMC level densities of these three
nuclei  are seen to be quite different from each other. Indeed we find that
$\Delta$ is positive for $^{55}$Co, close to zero for $^{55}$Fe and
negative  for $^{55}$Mn. The experimental level densities (dashed
lines)  are in good agreement with the Monte Carlo calculations,
suggesting a $T_z \equiv (N-Z)/2$ dependence of the level density, which is
usually  ignored in empirical formulae but is clearly observed in our
microsocpic calculations.

In conclusion, we have described a particle-reprojection method in the SMMC
that  enables the calculation of thermal properties for a series of nuclei
using  Monte Carlo sampling for a single nucleus. We applied the method
to the calculation of level densities.

This work was supported in part by the Department of Energy grant
No.\ DE-FG-0291-ER-40608, and by the Ministry of Education, Science, Sports
and  Culture of Japan (grants 08044056 and 11740137). Computational cycles
were
provided  by the Cornell Theory Center, the San Diego Supercomputer Center
and the NERSC high performance computing facility at LBL.

\end{document}